\documentclass
[superscriptaddress,secnumarabic,amssymb,amsmath,nobibnotes,aps,prd,showkeys,nofootinbib,nopacsnumber,twocolumn,10pt]{revtex4}%
\usepackage{graphicx}
\usepackage{bm}
\usepackage{xcolor}
\usepackage{amsmath}
\usepackage{amsfonts}
\usepackage[english]{babel}
\usepackage{amssymb}
\usepackage[utf8]{inputenc}%
\providecommand{\U}[1]{\protect\rule{.1in}{.1in}}

\newcommand{\be}{\begin{equation}}
\newcommand{\ee}{\end{equation}}

\newcommand{\mincir}{\raise
-3.truept\hbox{\rlap{\hbox{$\sim$}}\raise4.truept\hbox{$<$}\ }}
\newcommand{\magcir}{\raise
-3.truept\hbox{\rlap{\hbox{$\sim$}}\raise4.truept\hbox{$>$}\ }}

\begin{document}
\title{Indications of a Late-Time Transition to a Strongly Interacting Dark Sector}
\author{Andronikos Paliathanasis}
\email{anpaliat@phys.uoa.gr}
\affiliation{Institute of Systems Science, Durban University of Technology, Durban 4000,
South Africa}
\affiliation{Centre for Space Research, North-West University, Potchefstroom 2520, South Africa}
\affiliation{Departamento de Matem\`{a}ticas, Universidad Cat\`{o}lica del Norte, Avda.
Angamos 0610, Casilla 1280 Antofagasta, Chile}
\affiliation{National Institute for Theoretical and Computational Sciences (NITheCS), South Africa.}

\begin{abstract}
We explore the transition from the $\Lambda$CDM to an interacting dark sector
by introducing a model with a redshift threshold that controls the
onset of the energy transfer between the dark energy and the dark matter. Below the transition redshift, the interaction
between dark matter and dark energy becomes active, while at earlier times the
cosmological evolution coincides with that of $\Lambda$CDM. This approach
allows us to determine the epoch in the comic history where the interacting
effects have an impact in the description of the dark sector. We constrain the
free parameters of the model using late-time cosmological observations, namely
Cosmic Chronometers, DESI DR2 Baryonic Acoustic Oscillations, and Supernova
data from the Pantheon Plus, Union3.0, and DES-Dovekie catalogues. The
analysis provides an indication of a strong interacting term that describes
energy transfer from dark energy to dark matter, which is activated at low
redshifts. The PantheonPlus sample provides a threshold of $z_{T}<0.624$, the
Union3.0 sample yields $z_{T}=0.400_{-0.23}^{+0.021}$, and the DES-Dovekie
sample gives $z_{T}=0.371_{-0.26}^{+0.028}$. The model fits the data in a
similar way to the CPL parametrization, without the dark energy to cross the
phantom divide line.

\end{abstract}
\date{\today}
\maketitle

\section{Introduction}

The second data release of Baryon Acoustic Oscillation (BAO) measurements from
the Dark Energy Spectroscopic Instrument (DESI DR2) \cite{DESI:2025zpo,DESI:2025zgx,DESI:2025fii} provides support for
cosmological scenarios in which dark energy has a dynamical character \cite{Arora:2025msq,Kessler:2025kju,Zhang:2025bmk,Paliathanasis:2025cuc,Nagpal:2025omq,Ormondroyd:2025iaf,You:2025uon,Scherer:2025esj,Paradiso:2024pcb}. These findings are consistent with earlier indications obtained from previous data sets
\cite{DESI:2024mwx,Park:2024vrw,Park:2025azv,Park:2024pew,Alfano:2025gie,Carloni:2024zpl}. 

The origin and nature of the dark energy is unknown; nevertheless there are various approaches in the literature which investigate
the dynamical behaviour of the dark energy, see for instance \cite{Alestas:2025syk,SanchezLopez:2025uzw,
Pourtsidou:2025sdd,Yang:2025mws,Lu:2025sjg,Feng:2025cwi,Karmakar:2025yng,Samaddar:2025bjg,
Philcox:2025faf,Paliathanasis:2025hjw,Odintsov:2025jfq,Plaza:2025gcv,
DOnofrio:2025cuk,Najera:2025htf,Hossain:2025grx,Shajib:2025tpd,Anchordoqui:2025fgz,
Paliathanasis:2025xxm,Paliathanasis:2025mvy,Cai:2025mas,Ye:2025pem,Toomey:2025yuy,Zheng:2025vyv,
Shen:2025cjm,Hussain:2025uye,Yadav:2025vpx,Paliathanasis:2025dcr,
Paliathanasis:2025kmg,Yadav:2025wbc,Luciano:2025elo,Luongo:2024fww,quntomdesidr2} and references therein.

Interacting dark sector models have received high attention recently \cite{Wang:2016lxa,Farrar:2003uw,Lucca:2021dxo,Lucca:2020zjb,vanderWesthuizen:2025vcb,vanderWesthuizen:2025mnw,vanderWesthuizen:2025rip,Giare:2024smz,Giare:2024ytc,Benisty:2024lmj,Okengo:2024mub, Valiviita:2008iv, Caldera-Cabral:2008yyo,Zhai:2023yny,Pan:2019gop,Pan:2020zza,Yang:2019vni,Paliathanasis:2019hbi,Yang:2018qec}. In
such models, dark energy and dark matter form the dark sector of the universe.
While the total dark sector is conserved, the individual components are not,
indicating the presence of an energy exchange between the elements of the dark sector. In interacting models dark energy and dark matter can be seen as two
representations of the total cosmic fluid, and the energy transfer is analogue
to a phase transition. 

Cosmological models with dark energy-dark matter interaction have shown that
they can describe the recent BAO observations \cite{Zhu:2025lrk,Zhang:2025dwu,Li:2025muv,
Petri:2025swg,Li:2025ula, Pan:2025qwy,Silva:2025hxw, Shah:2025ayl,
Feng:2025mlo, Yang:2025uyv,vanderWesthuizen:2025iam}, since they can provide an
effective dark energy component with a dynamical character. Moreover, as
discussed recently in \cite{Guedezounme:2025wav}, interacting dark energy models can explain
the recent observation data without a phantom crossing scenario. 

In the literature the proposed interacting models have been considered the interaction to exist for the entire cosmic history. Recently, in \cite{Zhai:2025hfi} it was found that late-time observational data support energy transfer
within the dark sector. It was found that the introduction of
the BAO data set combined with the Planck 2018 Cosmic Microwave Background
(CMB) \cite{Planck:2018vyg} measurements and Supernova data of the PantheonPlus \cite{Brout:2022vxf} sample give a support to the interacting models. Thus, it is worth investigating in which period the late-time data interacting models are important for the description of the cosmos. 

In this work, we adopt a different approach. We consider late-time cosmological observations and introduce an interacting dark sector model characterized by a step-like transition from a non-interacting regime to an interacting one. Specifically, we introduce a redshift-dependent coupling parameter for the interaction, with a redshift threshold parameter $z_{T}$. For $z<z_{T}$ the interaction starts and supports a dynamical behaviour of dark energy, while for $z>z_{T}$ the cosmological model is non-interacting. With this consideration, we are able to identify which period of cosmological history supports a transition to an interacting scenario. Furthermore, we are able to investigate whether the interaction in the dark sector is strong or weak. 

We consider Supernova data, Cosmic Chronometers, and the BAO catalogue of DESI DR2, and we determine the transition redshift. We find that the data at redshifts $z<1$ support a strong interacting scenario with energy transfer from the dark energy to the dark matter. As a result, the model provides a larger value of the deceleration parameter at the present time, which indicates a slower expansion of the universe. Furthermore, we are able to describe the recent data without phantom crossing.
The structure of the paper is as follows.

In Section \ref{sec2} we introduce a
Friedmann--Lema\^{\i}tre--Robertson--Walker (FLRW) cosmological model with
interacting dark sector. We introduce an interacting function which is linear
to the energy density of the dark matter, where the coupling parameter that
quantifies the strength of the interaction we assume that it is redshift
dependent with a step transition. In Section \ref{sec3} we discuss the method
that we applied in this work for the statistical analysis of the theoretical
models with the observational data. The results from the observational
constraints are presented in Section \ref{sec4}, where we compare our model
with the $\Lambda$CDM, the CPL and the interacting model with a constant
coupling parameter. Finally, in Section \ref{sec5} we summarize our results
and draw our conclusions.

\section{Cosmological Interacting Models}

\label{sec2}

We consider a homogeneous and isotropic universe described by the spatially
flat FLRW spacetime%
\begin{equation}
ds^{2}=-dt^{2}+a^{2}(t)\left(  dx^{2}+dy^{2}+dz^{2}\right)  \;. \label{ln.01}%
\end{equation}
where $a\left(  t\right)  $ is the radius of a three-dimensional space. 

Within
the framework of\ General Relativity, the gravitational field is described by
the Einstein's field equations%
\begin{equation}
R_{\mu\nu}-\frac{R}{2}g_{\mu\nu}=T_{\mu\nu}, \label{ln.02}%
\end{equation}
where $T_{\mu\nu}$ describes the cosmic fluid which inherits the symmetries of
the background space. 

In the $1+3$ decomposition the energy momentum
tensor $T_{\mu\nu}$ is defined as%
\begin{equation}
T_{\mu\nu}=\left(  \rho+p\right)  u_{\mu}u_{\nu}+pg_{\mu\nu}\;, \label{ln.03}%
\end{equation}
with $\rho$ to be the energy density, $p=p\left(  t\right)  $ the pressure
component. $u^{\mu}$ is the fluid velocity, which we consider to be the
comoving, $u^{\mu}=\delta_{t}^{\mu}$, $u^{\mu}u_{\mu}=-1$, the
expansion rate $\theta\left(  t\right)  $ to be given by the expression
$\theta=3H$, where $H=\frac{\dot{a}}{a}$ is the Hubble function.

For the line element (\ref{ln.01}) the gravitational field equations read%
\begin{align}
3H^{2}  &  =\rho,\label{ln.04}\\
-2\dot{H}-3H^{2}  &  =p. \label{ln.05}%
\end{align}
The effective equation of state parameter for the cosmological fluid is
defined as $w_{eff}=\frac{p}{\rho}$, that is,
\begin{equation}
w_{eff}=-1-\frac{2}{3}\frac{\dot{H}}{H^{2}}. \label{ln.06}%
\end{equation}
Hence, cosmic acceleration is described when $w_{eff}<-\frac{1}{3}$.

Moreover, from the Bianchi identity we recover the conservation equation for
the cosmic fluid $T_{~~;\nu}^{\mu\nu}=0,$ that is,%
\begin{equation}
\dot{\rho}+3H\left(  \rho+p\right)  =0. \label{ln.08}%
\end{equation}

We assume that the energy momentum tensor $T_{\mu\nu}$ consists of three fluid components, the cold dark matter
\begin{equation}
T_{\mu\nu}^{m}=\rho_{m}u_{\mu}u_{\nu}, \label{ln.09}%
\end{equation}
the dark energy component
\begin{equation}
T_{\mu\nu}^{DE}=\left(  \rho_{d}+p_{d}\right)  u_{\mu}u_{\nu}+p_{d}g_{\mu\nu
}.~ \label{ln.10}%
\end{equation}
with equation of state parameter$~w_{d}=\frac{p_{d}}{\rho_{d}}$, and the
baryons
\begin{equation}
T_{\mu\nu}^{b}=\rho_{b}u_{\mu}u_{\nu},
\end{equation}
where we the baryons do not interact with the rest of the fluid components

Therefore, $\rho=\rho_{m}+\rho_{b}+\rho_{d}$ and $p=p_{d}.~$By replacing in
(\ref{ln.08}) it follows%
\begin{equation}
\left(  \rho_{m}+\rho_{b}+\rho_{d}\right)  ^{\cdot}+3H\left(  \rho_{m}%
+\rho_{b}+\rho_{d}+p_{d}\right)  =0,
\end{equation}
\bigskip or equivalently,~%
\begin{align}
\dot{\rho}_{b}+3H\rho_{b}  &  =0,\\
\dot{\rho}_{m}+3H\rho_{m}  &  =Q,\\
\dot{\rho}_{d}+3H\left(  \rho_{d}+p_{d}\right)   &  =-Q.
\end{align}

Here $Q\left(  t\right)  $ denotes the interaction term that governs the energy exchange between dark matter and dark energy. When the two fluid do not interact $Q\left(  t\right)  =0$. However, in interacting
scenarios$~Q\left(  t\right)  \neq0$, and when $Q\left(  t\right)  >0$, there
is energy transfer from dark energy to dark matter, while for $Q\left(
t\right)  <0$ dark matter is converted into dark energy.

There is a plethora of interacting models proposed in the literature \cite{Koshelev:2010umw,Mangano:2002gg,Duniya:2015nva,Majerotto:2009np,Majerotto:2009zz,Paliathanasis:2024abl}, for a review we refer the reader in \cite{vanderWesthuizen:2025vcb,vanderWesthuizen:2025mnw,vanderWesthuizen:2025rip}. 

In this
work we focus on the model
\[
Q_{A}=\alpha\left(  z\right)  H\rho_{m},
\]
where $\alpha\left(  z\right)  $ is the coupling parameter that
quantifies the strength of the interaction between dark energy and dark
matter. This parameter is usually considered constant,~$\alpha\left(
z\right)  =\alpha$. However, in this study we allow $\alpha\left(  z\right)  $
to vary with redshift, introducing a step transition,
\begin{equation}
\alpha\left(  z\right)  =\frac{\alpha}{2}\left(  1-\tanh\left(  \delta\left(
z-z_{T}\right)  \right)  \right)
\end{equation}
where parameter $z_{T}$ indicates the redshift threshold for the onset of interaction
and $\delta$ controls the sharpness of the step. We consider $z_{T}>0$ and
$\delta>>1$ in order the transition to the interacting sector to be fast.  

Interacting models with varying coupling parameters were introduced before in \cite{Yang:2019uzo} and applied recently for the study of the DESI DR2 data in \cite{Yang:2025uyv}. Moreover,  model where the coupling parameter change sing proposed in \cite{Silva:2025hxw}

The existence of the step transition allows us to understand for which
redshifts the interacting models become important, as also allow us to have a
cosmological theory where for $z>z_{T}$ the theory behaves as a non
interacting model.

In the following, we focus on the case where dark energy is described by a
cosmological constant, $w_{d}=-1$. For $z<z_{T},~$once the interaction is
triggered, we examine deviations from $\Lambda$CDM.

\section{Data \&\ Methodology}

\label{sec3}

In the following lines, we briefly discuss the cosmological data employed in
this work and the tools that we follow for the analysis of the constraints.

\subsection{Cosmological Data}

We focus on cosmological observations with of the late-time, that is, with
redshift $z<2.5$. In particular we consider the Supernova data, the Cosmic
Chronometers and the Baryonic Acoustic Oscillations.

\subsubsection{Supernova}

We consider three different samples for the Supernova (SNIa) data, the
PantheonPlus (PP) \cite{Brout:2022vxf} the Union3.0 (U3) \cite{rubin2023union}
and the DES-Dovekie (DESD) \cite{DES:2025sig} catalogues. These catalogues
provide the observable distance modulus\ $\mu^{obs}$ at the redshift $z$.
The PP sample contains 1,701 light curves corresponding to 1,550
spectroscopically supernovae events within the redshifts $10^{-3}<z<2.27$. In
this work we consider the PP sample without the SH0ES Cepheid calibration. 

On the other hand, the U3 catalogue\ includes 2,087 supernova events in the same
redshift range as PP, with 1,363 common events with the PP. However, the
analysis of the photometric observations is different between the two
catalogues. 

The DESD catalogue is most recent set release of supernova data
with 1820 supernova events with redshifts $z<1.13$ \cite{DES:2025sig}. The
catalogue released after the re-analysis of the five years data of the Dark
Energy Survey \ of Type Ia supernova (DES-SN5YR), using an improved
photometric calibration approach. 

In a spatially flat FLRW geometry, the
theoretical value of the distance modulus $\mu^{th}$ follows from the relation
$\mu^{{th}}=5\log_{10}D_{L}+25$, where $D_{L}=c\,(1+z)\int_{0}^{z}%
\frac{dz^{\prime}}{H(z^{\prime})}$ is the luminosity distance, and $c$ is the
speed of light.

\subsubsection{Cosmic Chronometers}

The cosmic chronometers are passively evolving galaxies that provide us with
direct measure of the Hubble parameter. They are model independent direct
measures of the Hubble parameter, leading to the Observable Hubble Dataset
(OHD). 

In this study we employ the 31 data presented in
\cite{moresco2020setting}, supplemented with three additional measurements
from the analysis of the DESI DR1 observations \cite{Loubser:2025fzl}.

\subsubsection{Baryonic Acoustic Oscillations}

We consider the recent release of the Dark Energy Spectroscopic Instrument
(DESI DR2) baryon acoustic oscillation (BAO)
observations~\cite{DESI:2025zpo,DESI:2025zgx,DESI:2025fii}, which provides
measurements of the transverse comoving angular distance ratio, $\frac{D_{M}%
}{r_{drag}}=\frac{D_{L}}{\left(  1+z\right)  r_{drag}},~$the volume-averaged
distance ratio, $\frac{D_{V}}{r_{drag}}=\frac{\left(  zD_{H}D_{M}^{2}\right)
^{1/3}}{r_{drag}}$ and the and the Hubble distance ratio $\frac{D_{H}}{r_{d}%
}=\frac{c}{r_{drag}H(z)},~$at seven distinct redshifts, where $~r_{drag}$.

\subsection{Bayesian Analysis}

For the statistical analysis we make use of the Bayesian inference framework
COBAYA\footnote{https://cobaya.readthedocs.io/}~\cite{cob1,cob2}, with the use
of a custom theory together with the MCMC sampler~\cite{mcmc1,mcmc2}. 

For each test we consider the following combination of data SNIa\&OHD\&BAO, where SNIa
refers to the PP, U3 or DESD catalogues. 

We analyze the chains of the MCMC
sampler with the GetDist library\footnote{https://getdist.readthedocs.io/}%
~\cite{getd}, and we determine the posterior variables which maximize the
likelihood $\mathcal{L}_{\max}=\exp\left(  -\frac{1}{2}\chi_{\min}^{2}\right)
$, that is
\[
\mathcal{L}_{\max}=\mathcal{L}_{SNIa}\times\mathcal{L}_{OHD}\times
\mathcal{L}_{BAO},
\]
that is%
\[
\chi_{\min}^{2}=\chi_{\min\left(  SNIa\right)  }^{2}+\chi_{\min\left(
OHD\right)  }^{2}+\chi_{\min\left(  BAO\right)  }^{2}.
\]

\subsubsection{Model Comparison}

We perform the analysis for the four-different models, $\Lambda$CDM, the
Chevallier-Polarski-Linder (CPL) parametrization ($w_{0}w_{a}$CDM),
interacting model without threshold, and interacting model with the threshold
These models have different number of degrees of freedom, and in order to make
a statistical comparison we apply the Akaike Information Criterion
(AIC)~\cite{AIC}. For a large value of datasets. 

The AIC parameter is
defined through the algebraic relation
\begin{equation}
{AIC}\simeq\chi_{\min}^{2}+2\kappa,
\end{equation}
where $\kappa$ refers to the dimension space of the free parameters for each model.

For the $\Lambda$CDM the free parameters are $\left\{  H_{0},\Omega
_{m0},r_{drag}\right\}  ,$that is $\kappa_{\Lambda\text{CDM}}=3$, for the CPL
parametric dark energy model, the free parameters are five, that is,
$\kappa_{w_{0}w_{a}CDM}=5$, $\left\{  H_{0},\Omega_{m0},r_{drag},w_{0}%
,w_{a}\right\}  $, and for the interacting model of our analysis without
threshold $\left(  Q_{A}^{0}\right)  $ $\kappa_{I}=4$, and free parameters are
$\left\{  H_{0},\Omega_{m0},r_{drag},\alpha\right\}  ,$ while for the
interacting models with threshold ($Q_{A}$), $\kappa_{IT}=5$, where $\left\{
H_{0},\Omega_{m0},r_{drag},\alpha,z_{T}\right\}  $. As far as the parameter
$\delta$ is concerned, we assume $\delta=10^{4}$ such that the transition to the
interacting era to be sharp. For the baryons we consider the value obtained by by the Planck 2018 mission \cite{Planck:2018vyg}

Akaike's scale provides information on which model offers a better fit to the
data based on the value of $\Delta AIC=AIC_{A}-AIC_{B}.$ The Akaike scale
explicitly penalizes models with a large number of free parameters, thus
reducing overfitting. Specifically, for $\lvert\Delta{AIC}\rvert<2$, the
models are equally consistent with the data, when $2<\lvert\Delta{AIC}%
\rvert<6$, there is weak evidence in favor of the model with the smaller AIC
value; if $6<\lvert\Delta{AIC}\rvert<10$, the evidence is strong; and for
$\lvert\Delta{AIC}\rvert>10$, there is a clear evidence favoring for the model
with the lower AIC.%

\begin{table*}[tbp] \centering
\caption{Posterior variables for the interacting models as given by the analysis of the MCMC chains.}%
\begin{tabular}
[c]{cccccc}\hline\hline
\textbf{Data Set} & $\mathbf{H}_{0}$ & $\mathbf{\Omega}_{m0}$ & $\mathbf{r}%
_{drag}$ & $\alpha$ & $z_{T}$\\\hline
\multicolumn{6}{c}{\textbf{Model }$\mathbf{Q}_{A}$}\\\hline
\textbf{PP\&OHD\&BAO} & $68.2_{-1.7}^{+1.7}$ & $0.334_{-0.074}^{+0.029}$ &
$146.9_{-3.6}^{+3.1}$ & $1.55_{-1.4}^{+0.55}$ & \thinspace$<0.624$\\
\textbf{U3\&OHD\&BAO} & $66.6_{-1.8}^{+1.8}$ & $0.490_{-0.15}^{+0.10}$ &
$147.1_{-3.4}^{+3.4}$ & $2.4_{-1.9}^{1.1}$ & $0.400_{-0.23}^{+0.021}$\\
\textbf{DD\&OHD\&BAO} & $67.9_{-1.6}^{+1.6}$ & $0.406_{-0.11}^{+0.071}$ &
$146.8_{-3.4}^{+3.4}$ & $2.2_{-2.2}^{+1.0}$ & $0.371_{-0.26}^{+0.028}$\\\hline
\multicolumn{6}{c}{\textbf{Model }$\mathbf{Q}_{A}^{0}$}\\\hline
\textbf{PP\&OHD\&BAO} & $68.0_{-1.6}^{+1.9}$ & $0.300_{-0.035}^{+0.009}$ &
$147.2_{-4.1}^{+3.1}$ & $0.203_{-0.18}^{+0.021}$ & $-$\\
\textbf{U3\&OHD\&BAO} & $67.1_{-1.6}^{+2.4}$ & $0.337_{-0.066}^{+0.002}$ &
$147.7_{-4.2}^{+3.2}$ & $0.324_{-0.25}^{-0.011}$ & $-$\\
\textbf{DD\&OHD\&BAO} & $67.9_{-1.6}^{+1.9}$ & $0.301_{-0.034}^{+0.007}$ &
$147.6_{-4.3}^{+3.0}$ & $0.229_{-0.19}^{+0.015}$ & $-$\\\hline\hline
\end{tabular}
\label{data1}%
\end{table*}%

\section{Observational Constraints}

\label{sec4}

We continue our discussion with the presentation of the analysis of the MCMC
chains, for the three different combinations of the datasets. The posterior
variables for the interacting models $Q_{A},~Q_{A}^{0},$ with and without the
initiate mechanism are presented in Table \ref{data1}. Moreover, in Tables
\ref{data2} and \ref{data3} we provide the statistical comparison of the two
interacting models $Q_{A}$ and$~Q_{A}^{0}$, with the $\Lambda$CDM and the CPL
models.

\begin{figure}[h]
\centering\includegraphics[width=0.52\textwidth]{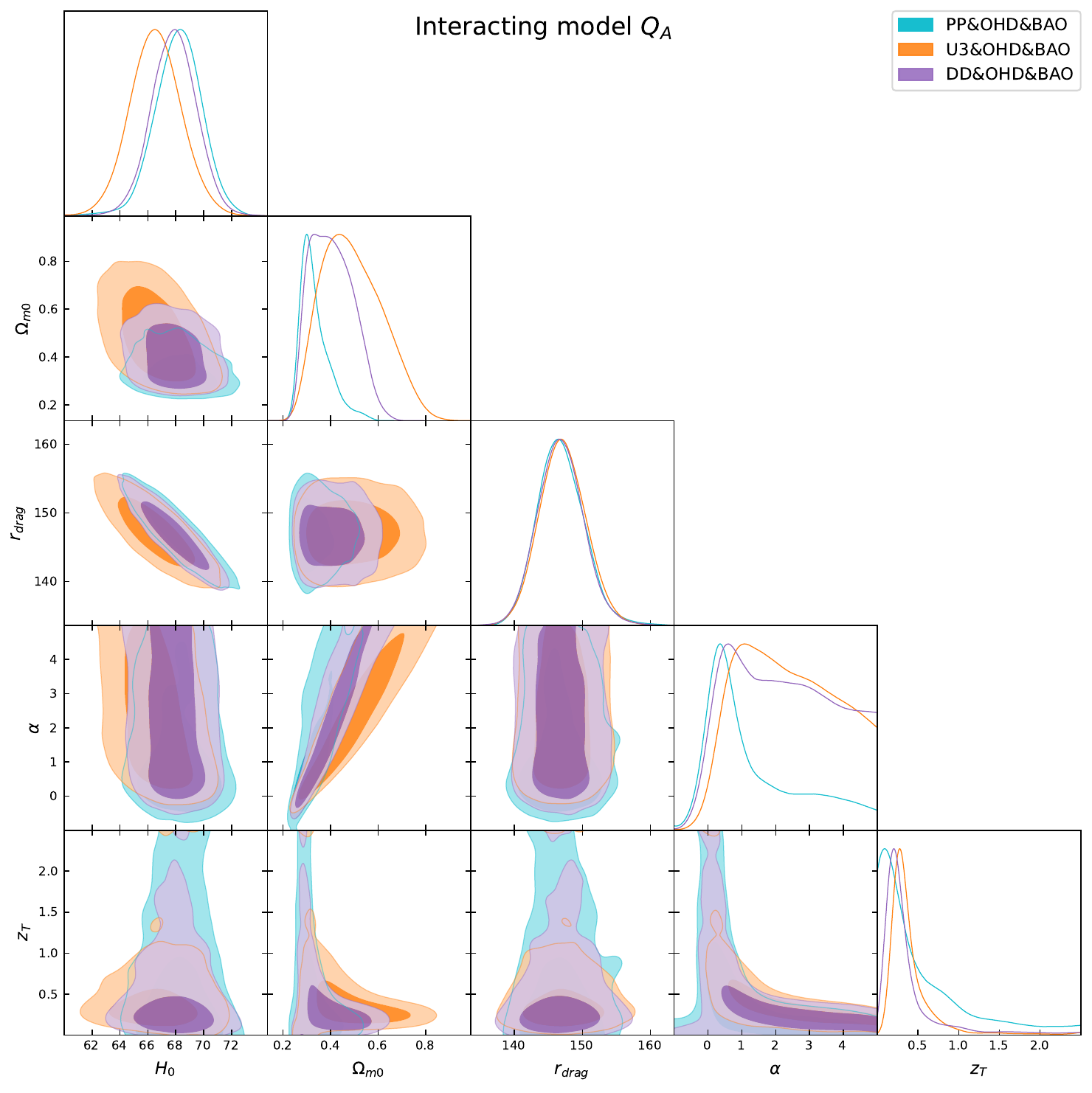}\caption{Confidence
space for the posterior parameters $\left\{  H_{0},\Omega_{m0},r_{drag}%
,\alpha,z_{T}\right\}  $ for the interacting model $Q_{A}$. }%
\label{fig1}
\end{figure}

\begin{figure}[h]
\centering\includegraphics[width=0.5\textwidth]{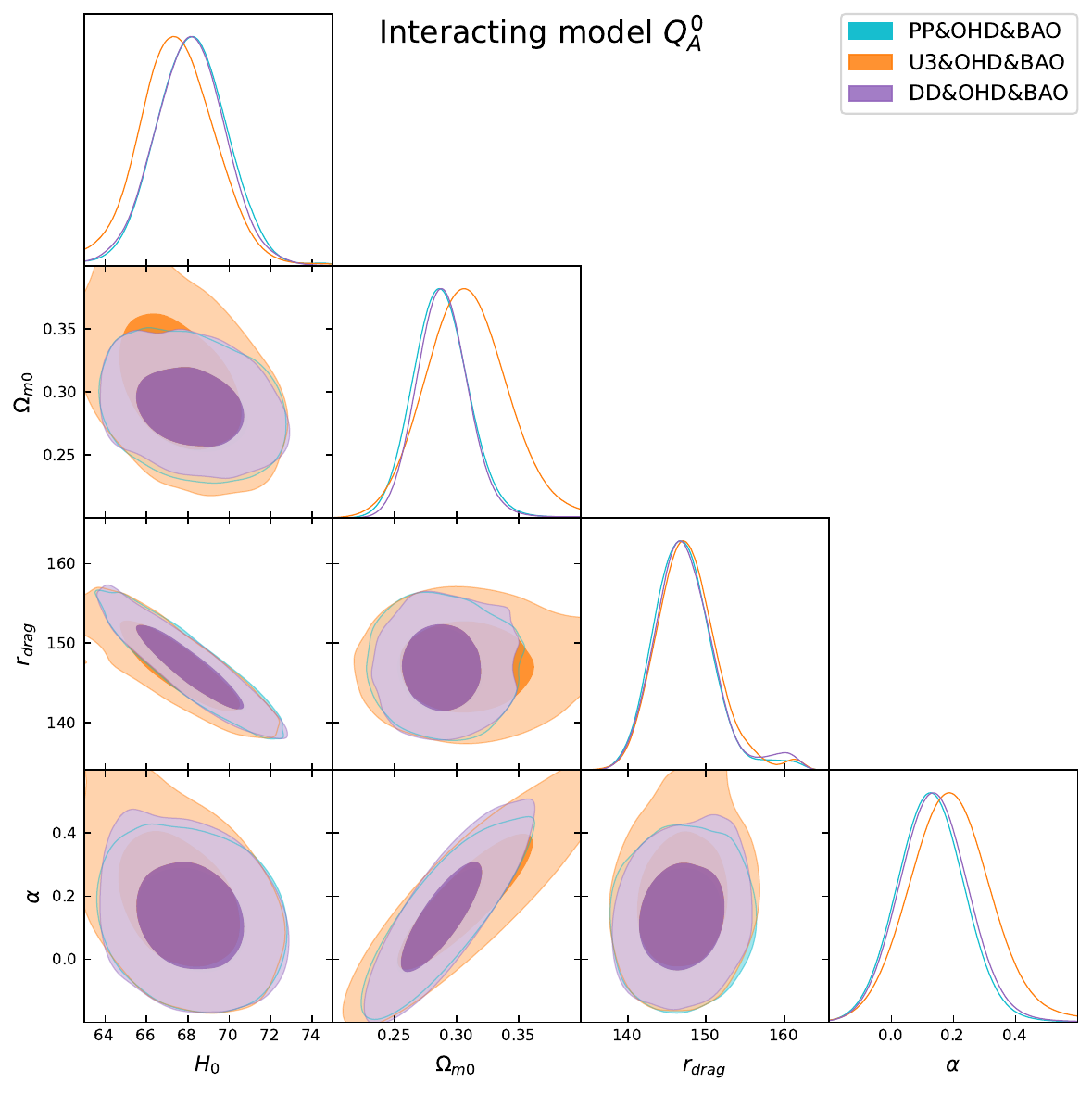}\caption{Confidence
space for the posterior parameters $\left\{  H_{0},\Omega_{m0},r_{drag}%
,\alpha_{0}\right\}  ~$for the interacting model $Q_{A}^{0}$. }%
\label{fig2}
\end{figure}

In Figs. \ref{fig1} and \ref{fig2} we present the contours for the
confidence space of the free parameters for the $Q_{A}$ and $Q_{A}^{0}$ models respectively.%

\begin{table*}[tbp] \centering
\caption{Statistical comparison of the interacting model with with the initiate mechanism with respect to the $\Lambda$CDM and the CPL.}%
\begin{tabular}
[c]{cccccc}\hline\hline
\textbf{Data Set} & \textbf{Reference Model} & \textbf{Base Model} &
$\mathbf{\Delta\chi}_{\min}^{2}$ & $\mathbf{\Delta AIC}$ & \textbf{Akaike's
Scale}\\\hline
\textbf{PP\&OHD\&BAO} & $\Lambda$CDM & $Q_{A}$ & $-3.37$ & $+0.63$ &
Inconclusive\\
\textbf{PP\&OHD\&BAO} & CPL & $Q_{A}$ & $+0.01$ & $+0.01$ & Inconclusive\\
\textbf{U3\&OHD\&BAO} & $\Lambda$CDM & $Q_{A}$ & $-7.9$ & $-3.9$ & Weak
Evidence in favor of $Q_{A}$\\
\textbf{U3\&OHD\&BAO} & CPL & $Q_{A}$ & $+0.34$ & $+0.34$ & Inconclusive\\
\textbf{DD\&OHD\&BAO} & $\Lambda$CDM & $Q_{A}$ & $-6.42$ & $-2.42$ & Weak
Evidence in favor of $Q_{A}$\\
\textbf{DD\&OHD\&BAO} & CPL & $Q_{A}$ & $-1.65$ & $-1.65$ &
Inconclusive\\\hline\hline
\end{tabular}
\label{data2}%
\end{table*}%
%

\begin{table*}[tbp] \centering
\caption{Statistical comparsion of the interacting model without the initiate mechanism with respect to the $\Lambda$CDM and the CPL.}%
\begin{tabular}
[c]{cccccc}\hline\hline
\textbf{Data Set} & \textbf{Reference Model} & \textbf{Base Model} &
$\mathbf{\Delta\chi}_{\min}^{2}$ & $\mathbf{\Delta AIC}$ & \textbf{Akaike's
Scale}\\\hline
\textbf{PP\&OHD\&BAO} & $\Lambda$CDM & $Q_{A}^{0}$ & $-1.86$ & $+0.14$ &
Inconclusive\\
\textbf{PP\&OHD\&BAO} & CPL & $Q_{A}^{0}$ & $+1.52$ & $-0.48$ & Inconclusive\\
\textbf{U3\&OHD\&BAO} & $\Lambda$CDM & $Q_{A}^{0}$ & $-2.7$ & $+0.7$ &
Inconclusive\\
\textbf{U3\&OHD\&BAO} & CPL & $Q_{A}^{0}$ & $+5.54$ & $+3.54$ & Weak Evidence
in favor of CPL\\
\textbf{DD\&OHD\&BAO} & $\Lambda$CDM & $Q_{A}^{0}$ & $-2.2$ & $-0.2$ &
Inconclusive\\
\textbf{DD\&OHD\&BAO} & CPL & $Q_{A}^{0}$ & $+2.57$ & $-0.57$ &
Inconclusive\\\hline\hline
\end{tabular}
\label{data3}%
\end{table*}%

\subsection{\textbf{PP\&OHD\&BAO}}

For the first combination data, where we consider the PP catalogue for the
SNIa, from the interacting model $Q_{A}$ we find the posterior variables
$H_{0}=68.2_{-1.7}^{+1.7},~\Omega_{m0}=0.334_{-0.074}^{+0.029}$,
$r_{drag}=146.9_{-3.6}^{+3.1}$, $\alpha=1.55_{-1.4}^{+0.55}$ and $z_{T}%
<0.624$, while for the interaction without the threshold, i.e. $Q_{A}^{0}$, we
derive $H_{0}=68.0_{-1.6}^{+1.9},~\Omega_{m0}=0.300_{-0.035}^{+0.009}$,
$r_{drag}=147.2_{-4.1}^{+3.1}$, $\alpha=0.203_{-0.18}^{+0.021}$. \ 

The comparison of the statistical parameters shows that model $Q_{A}$ and CPL
fit the data with the same way, $\chi_{\min}^{2}\left(  Q_{A}\right)
-\chi_{\min}^{2}\left(  CPL\right)  \simeq+0.01$ and better in comparison to
the $\Lambda$CDM and $Q_{A}^{0}$ models,~that is, $\chi_{\min}^{2}\left(
Q_{A}\right)  -\chi_{\min}^{2}\left(  \Lambda\right)  \simeq-3.37$,
$\chi_{\min}^{2}\left(  Q_{A}^{0}\right)  -\chi_{\min}^{2}\left(
\Lambda\right)  \simeq-1.86$ and $\chi_{\min}^{2}\left(  Q_{A}^{0}\right)
-\chi_{\min}^{2}\left(  CPL\right)  \simeq+1.52$. However, due to the
different number of degrees of freedom, the four models are statistically equivalent.

\subsection{\textbf{U3\&OHD\&BAO}}

We consider the U3 catalogue for the SNIa. For the interacting model $Q_{A}$
we find the posterior variables $H_{0}=66.6_{-1.8}^{+1.8},~\Omega
_{m0}=0.490_{-0.15}^{+0.10}$, $r_{drag}=147.1_{-3.4}^{+3.4}$, $\alpha
=2.4_{-1.9}^{1.1}$ and $z_{T}=0.400_{-0.23}^{+0.021}$, while for $Q_{A}^{0}$
it follows $H_{0}=67.1_{-1.6}^{+2.4},~\Omega_{m0}=0.337_{-0.066}^{+0.002}$,
$r_{drag}=147.7_{-4.2}^{+3.2}$, $\alpha=0.324_{-0.25}^{-0.011}$.

From Tables \ref{data2} and \ref{data3} it follows $\chi_{\min}^{2}\left(
Q_{A}\right)  -\chi_{\min}^{2}\left(  CPL\right)  \simeq+0.34$, $\chi_{\min
}^{2}\left(  Q_{A}\right)  -\chi_{\min}^{2}\left(  \Lambda\right)  \simeq
-7.9$, $\chi_{\min}^{2}\left(  Q_{A}^{0}\right)  -\chi_{\min}^{2}\left(
CPL\right)  \simeq+5.54$ and $\chi_{\min}^{2}\left(  Q_{A}^{0}\right)
-\chi_{\min}^{2}\left(  \Lambda\right)  \simeq-2.7.$

Thus, $Q_{A}$ and the CPL model fit the data in the same way, while in
comparison to the $\Lambda$CDM and the $Q_{A}^{0}$, there is a weak evidence
in favor to the $Q_{A}$. \ However,

\subsection{\textbf{DD\&OHD\&BAO}}

Finally, the application of the DD sample for the SNIa, for model $Q_{A}$ we
derive the posterior variables $H_{0}=67.9_{-1.6}^{+1.6},~\Omega
_{m0}=0.406_{-0.11}^{+0.071}$, $r_{drag}=146.8_{-3.4}^{+3.4}$, $\alpha
=2.2_{-2.2}^{+1.0}$ and $z_{T}=0.371_{-0.26}^{+0.028}$, while for $Q_{A}^{0}$
we calculate $H_{0}=67.9_{-1.6}^{+1.9},~\Omega_{m0}=0.301_{-0.034}^{+0.007}$,
$r_{drag}=147.6_{-4.3}^{+3.0}$, $\alpha=0.229_{-0.19}^{+0.015}$.

Moreover, Tables \ref{data2} and \ref{data3} reveal, $\chi_{\min}^{2}\left(
Q_{A}\right)  -\chi_{\min}^{2}\left(  CPL\right)  \simeq-1.65$, $\chi_{\min
}^{2}\left(  Q_{A}\right)  -\chi_{\min}^{2}\left(  \Lambda\right)
\simeq-6.42$, $\chi_{\min}^{2}\left(  Q_{A}^{0}\right)  -\chi_{\min}%
^{2}\left(  CPL\right)  \simeq+2.57$ and $\chi_{\min}^{2}\left(  Q_{A}%
^{0}\right)  -\chi_{\min}^{2}\left(  \Lambda\right)  \simeq-2.2$.

Therefore, model $Q_{A}$ fits the data in a better way than the rest of the
models, and it comparison to the $\Lambda$CDM and the $Q_{A}^{0}$ the data
provide a weak evidence in favor to the $Q_{A}$. Nevertheless, Akaike's scale
indicate that $Q_{A}$ and CPL models are statistically equivalent.

\begin{figure*}[h]
\centering\includegraphics[width=0.8\textwidth]{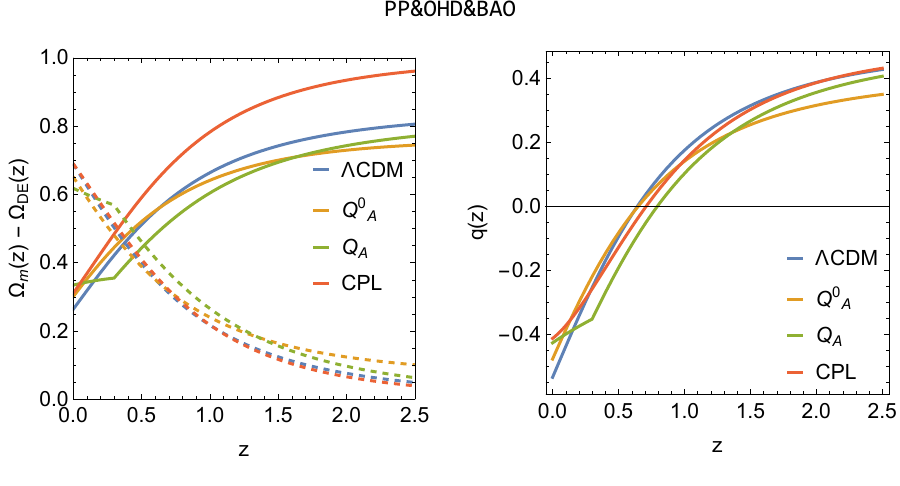}\caption{PP\&OHD\&BAO:\ Left
Fig: Qualitative evolution of the $\Omega_{m}\left(  z\right)  ~$(solid lines)
and~$\Omega_{DE}\left(  z\right)  $ (dashed lines) for the $\Lambda$CDM, the
$Q_{A}^{0}$ and $Q_{A}$ interacting models. Right Fig: Qualitative evolution
of the deceleration parameter $q\left(  z\right)  =\frac{1}{2}\left(
1+3w_{tot}\left(  z\right)  \right)  $. Plots are for the best-fit parameters.
Blue lines are for the $\Lambda$CDM, orange and green lines are for the
$Q_{A}^{0}$ and $Q_{A}$ interacting models respectively, red lines are for the
CPL parametrization.}%
\label{fig3}
\end{figure*}

\begin{figure*}[h]
\centering\includegraphics[width=0.8\textwidth]{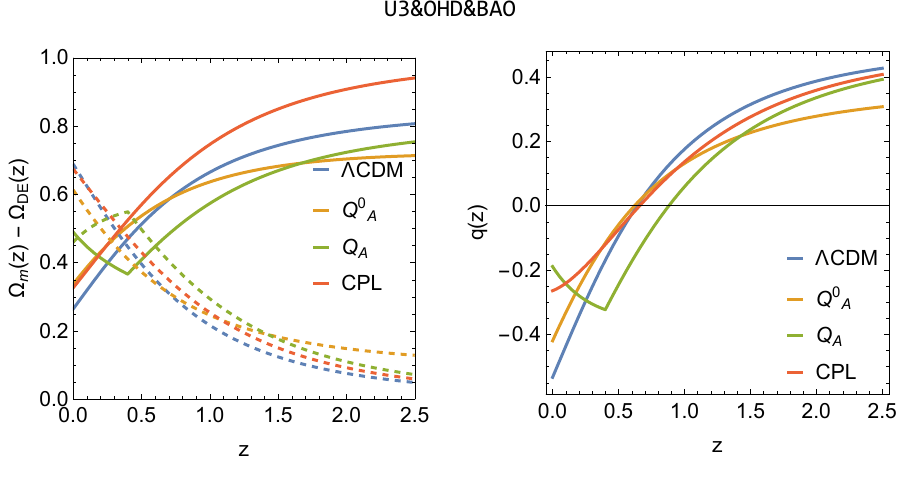}\caption{U3\&OHD\&BAO:\ Left
Fig: Qualitative evolution of the $\Omega_{m}\left(  z\right)  ~$(solid lines)
and~$\Omega_{DE}\left(  z\right)  $ (dashed lines) for the $\Lambda$CDM, the
$Q_{A}^{0}$ and $Q_{A}$ interacting models. Right Fig: Qualitative evolution
of the deceleration parameter $q\left(  z\right)  =\frac{1}{2}\left(
1+3w_{tot}\left(  z\right)  \right)  $. Plots are for the best-fit parameters.
Blue lines are for the $\Lambda$CDM, orange and green lines are for the
$Q_{A}^{0}$ and $Q_{A}$ interacting models respectively, red lines are for the
CPL parametrization.}%
\label{fig4}%
\end{figure*}

\begin{figure*}[h]
\centering\includegraphics[width=0.8\textwidth]{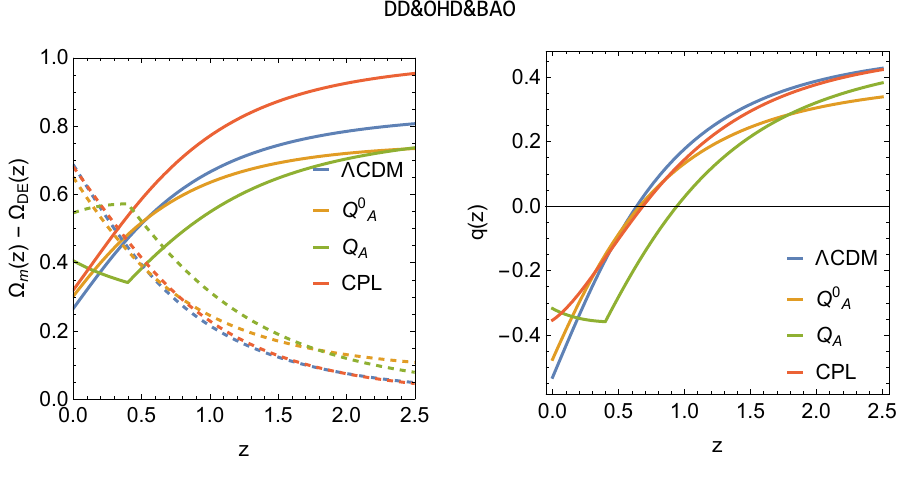}\caption{DD\&OHD\&BAO:\ Left
Fig: Qualitative evolution of the $\Omega_{m}\left(  z\right)  ~$(solid lines)
and~$\Omega_{DE}\left(  z\right)  $ (dashed lines) for the $\Lambda$CDM, the
$Q_{A}^{0}$ and $Q_{A}$ interacting models. Right Fig: Qualitative evolution
of the deceleration parameter $q\left(  z\right)  =\frac{1}{2}\left(
1+3w_{tot}\left(  z\right)  \right)  $. Plots are for the best-fit parameters.
Blue lines are for the $\Lambda$CDM, orange and green lines are for the
$Q_{A}^{0}$ and $Q_{A}$ interacting models respectively, red lines are for the
CPL parametrization.}%
\label{fig5}%
\end{figure*}

\subsection{Discussion}

The analysis of the observational constraints for the interacting model
$Q_{A}$ with the three different data sets, reveal that the transition to a an
interacting sector is at small redshifts $z<0.5$. Moreover, the
interacting coupling parameter $\alpha$ for the $Q_{A}$ is found to be
approximately eight times larger than the coupling parameter for the model
$Q_{A}$, from where we infer that the late-time observational data, with
$z<0.5$, support a strong interacting scenario.

In Figs. \ref{fig3}, \ref{fig4} and \ref{fig5} we present the qualitative
evolution of the energy densities for the dark matter $\Omega_{m}\left(
z\right)  $, the dark energy $\Omega_{DE}\left(  z\right)  $, and the
deceleration parameters $q(z)$ for the best-fit parameters given in Table \ref{data1}
for the three different data sets and we compare them with the $\Lambda$CDM and
the CPL parametrization.

Because of the strong positive interacting coupling parameter for model
$Q_{A}$, the data suggests a transition to a slower acceleration rate at the
present time in comparison to the $\Lambda$CDM, which is in agreement with the
behaviour provided by the CPL model. On the other hand, for $z>0.5$, the
interacting model $Q_{A}$ suggests a higher acceleration rate for the
universe, while for large values of redshift, due to the different values of
the $\Omega_{m}$ at the transition point $z_{T}$. Recall that for $z>z_{T},$
$\ Q_{A}$ and $\Lambda$CDM are identical.

\section{Conclusions}

\label{sec5}

Models with interaction in the dark sector of the universe provide a simple
mechanism for the description of the dynamical character of dark energy, and
it has been shown that they can challenge the $\Lambda$CDM model in the
description of late-time observational data. In this work, we considered a
phenomenological interacting dark sector with a redshift threshold for the
onset of the interacting term. This consideration allowed us to understand
during which period of cosmic evolution the interacting terms appear and play
an contribute in the phase-transition of the dark sector.

In particular, we considered an extension of the $\Lambda$CDM model with an
interacting term linear in the dark matter component and a step potential that
depends on the transition redshift $z_{T}$. For $z<z_{T}$ the interaction
appears, while for $z>z_{T}$ the model behaves as $\Lambda$CDM. We tested this
model using recent cosmological data, specifically cosmic chronometers, BAO
data, and SNIa data from the PP, U3, and DD samples. For the three different
combinations of data sets, we determined the posterior space of the free
parameters of the model under consideration. We performed the same tests for
the same interacting model without the onset mechanism, for $\Lambda$CDM, and
for the CPL parametrized dark energy model.

We found that the interacting model with the onset mechanism fits the data
better than the simple interacting model. The data show a weak preference for
our model compared to $\Lambda$CDM or the interacting model with a constant coupling parameter.
Nevertheless, our model is statistically equivalent to the CPL model. Our
interacting model and the CPL model share the same number of free parameters
and fit the three data sets in a similar way. Specifically, for PP\&OHD\&BAO
we find $\chi_{\min}^{2}\left(  Q_{A}\right)  -\chi_{\min}^{2}\left(
CPL\right)  \simeq+0.01$, for U3\&OHD\&BAO we derive $\chi_{\min}^{2}\left(
Q_{A}\right)  -\chi_{\min}^{2}\left(  CPL\right)  \simeq+0.34$, while for the
combined data set DD\&OHD\&BAO the analysis provides $\chi_{\min}^{2}\left(
Q_{A}\right)  -\chi_{\min}^{2}\left(  CPL\right)  \simeq-1.65$. Importantly, our interacting model is able to reproduce the observational data at the same level as the CPL model without requiring a phantom crossing of the dark energy equation of state.

Furthermore, we found that the transition point $z_{T}$ for the PP\&OHD\&BAO
data set is $z_{T}<0.624$, for U3\&OHD\&BAO the transition is at
$z_{T}=0.400_{-0.23}^{+0.021}$, and for DD\&OHD\&BAO the transition point is
at $z_{T}=0.371_{-0.26}^{+0.028}$. All three data sets favor a large and positive coupling parameter, indicating a strong energy transfer from dark energy to dark matter at late times. As a result,
the deceleration parameter for $z<z_{T}$ increases, such that the expansion of
the universe proceeds at a smaller rate. 

In this work we have not considered early universe observational data. However, from the results of \cite{Zhai:2025hfi} it is known that the late-time data support an interacting model, while early-data supports $\Lambda$CDM, and because for $z>z_{T}$ our model behaves identical with the $\Lambda$CDM we expect that our model fits the early-time data in a similar way to the $\Lambda$CDM.

The exact mechanism for the energy transfer between dark energy and dark
matter is not known. However, we can refer to the dark sector as a single
fluid with two different components. In this one fluid description, the energy
transfer mechanism is effective and can be interpreted as a phase transition
of the fluid. Within this framework, the onset mechanism that triggers the
phase transition can be easily introduced. Such behaviour is provided by the hameleon mechanism \cite{Khoury:2003rn} and in general within the symmetron cosmology \cite{Hinterbichler:2011ca}. 

In future work, we plan to
investigate the introduction of the step mechanism within a theoretical framework.

\begin{acknowledgments}
AP acknowledges the support from FONDECYT Grant 1240514 and from VRIDT through
Resoluci\'{o}n VRIDT No. 096/2022 and Resoluci\'{o}n VRIDT No. 098/2022.
\end{acknowledgments}

\bibliography{biblio}
\end{document}